\documentclass[11pt]{article}

\usepackage[utf8]{inputenc}
\usepackage[T1]{fontenc}
 
\usepackage{a4}
\usepackage{amsfonts,amsmath,mathrsfs,}
\usepackage{amssymb}
\usepackage{euscript}
\usepackage{epsfig}
\usepackage{color}
\usepackage[hyphens]{url}
\usepackage[hyphenbreaks]{breakurl}
\usepackage{tikz}
\usetikzlibrary{decorations.pathreplacing}
\usetikzlibrary{arrows}
\tikzset{>=triangle 45}

\newtheorem{theorem}{Theorem}
\newtheorem{definition}[theorem]{Definition}
\newtheorem{lemma}[theorem]{Lemma}

\newtheorem{proposition}[theorem]{Proposition}
\newtheorem{fact}[theorem]{Fact}

\newenvironment{proof}{\noindent {\it Proof~:}\ }{\ \rule{1mm}{2mm}\medskip}

\newcommand{\f}{\mathbb F}
\newcommand{\z}{\mathbb Z}

\renewcommand{\H}{\EuScript H}
\newcommand{\II}{{\mathbf I}}
\newcommand{\HH}{{\mathbf H}}
\newcommand{\GG}{{\mathbf G}}
\newcommand{\V}{\EuScript V}
\newcommand{\E}{\EuScript E}
\newcommand{\Z}{\EuScript Z}

\newcommand{\G}{\EuScript G}

\newcommand{\x}{{\mathbf x}}

\newcommand{\h}{{\mathbf h}}

\newcommand{\Co}{{\mathscr{C}}}
\newcommand{\CoX}{{\Co_X}}
\newcommand{\CoZ}{{\Co_Z}}
\newcommand{\CoXd}[2]{{\CoX(#1 \times #2)}}
\newcommand{\CoZd}[2]{{\CoZ(#1 \times #2)}}

\newcommand{\QCo}{{\EuScript{Q}}}
\newcommand{\QC}[2]{{\QCo(#1 \times #2)}}

\newcommand{\GXd}[2]{{#1 \times_X #2}}

\newcommand{\GZd}[2]{{#1 \times_Z #2}}
\newcommand{\dX}{{d_X}}
\newcommand{\dZ}{{d_Z}}
\newcommand{\dQ}{{d_{\QCo}}}
\newcommand{\kQ}{{k_{\QCo}}}
\newcommand{\Tan}[3]{{\EuScript{T}}(#1,#2,#3)}
\newcommand{\gprod}[2]{{#1 \times #2}}

\newcommand{\pc}{\mathbf H}
\newcommand{\eqdef}{\stackrel{\text{def}}{=}}
\newcommand{\up}[1]{\bar{#1}}
\newcommand{\supp}{{\text{supp}}}

\begin{document}

\title{Quantum LDPC codes with positive rate and minimum distance
proportional to $n^{1/2}$}

\author{Jean-Pierre Tillich\thanks{
INRIA, Projet Secret, BP 105, Domaine de Voluceau 
F-78153 Le Chesnay, France. 
Email: jean-pierre.tillich@inria.fr}
\and
Gilles Z\'emor\thanks{
Institut de Math\'ematiques de Bordeaux, UMR 5251,
Universit\'e Bordeaux 1,
351, cours de la Lib\'eration, F-33405 Talence Cedex, France 
Email: Gilles.Zemor@math.u-bordeaux1.fr}}

\date{January 11, 2013}

\maketitle

\begin{abstract}
The current best asymptotic lower bound on the minimum distance of
quantum LDPC codes with fixed non-zero rate is logarithmic in the
blocklength.
We propose a construction of quantum LDPC codes with fixed non-zero
rate and prove that the minimum distance grows proportionally to the
square root of the blocklength.
\end{abstract}

\let\thefootnote\relax\footnotetext{Material in this paper was
  presented in part at ISIT 2009, 799-803.}

\section{Introduction}
 LDPC codes~\cite{Gal63a} and their variants are one of the most
 satisfying answers to the problem of devising codes
 guaranteed by Shannon's theorem. 
 They display outstanding performance for a large class of error models with a 
fast decoding algorithm. Generalizing these codes to the quantum setting seems a promising way to 
devise powerful quantum error correcting codes for protecting,
 for instance, the 
very fragile superpositions manipulated in a quantum computer.
It should be emphasized that a fast decoding algorithm could be even more crucial in the quantum setting than in the classical one. In the classical case, when  error correction codes are used for communication over a noisy channel, the decoding time translates directly into communication delays. This has been the driving motivation to devise decoding schemes of low complexity, and is likely to be important in the quantum setting as well. However, there is an important additional motivation for efficient decoding in the quantum setting. Quantum computation is likely to require active stabilization. The decoding time thus translates into computation delays, and most importantly in error suppression delays. If errors accumulate faster than they can be identified, quantum computation may well become infeasible: fast decoding is an essential ingredient to fault-tolerant computation. 

 Besides the search for efficiently decodable codes, there is
  additional theoretical motivation for studying quantum LDPC codes that is
  totally absent from the classical setting. The capacity of the
  depolarizing channel, the quantum equivalent of the binary symmetric
channel, is unknown. What is clear however, is that capacity can only
be achieved by making use of degeneracy: the same syndrome must be
able to correct different error patterns. For this to be possible, the
stabilizer group must contain elements of reasonably small weight. In
other words, the quantum code must be at least partially
low-density.  
This intuition is confirmed by the concatenated construction \cite{DSS98a} (and the subsequent improvements
\cite{SS07a,FW08a}) improving upon the hashing lower bound on the capacity of the depolarizing channel.
All these constructions have many elements in the stabilizer group of very low Hamming weight. 
Quantum LDPC codes are not just about fast
decoding schemes, but may be the way towards a better understanding of
quantum channels from a purely information theory viewpoint.

Quantum 
generalizations of LDPC codes have been proposed in
\cite{MMM04a}. 
However, it has turned out that the design of high performance quantum
LDPC codes is much more complicated than in the classical setting. 
This is due to several reasons, the most obvious of which being that
the parity-check matrix of quantum LDPC codes must satisfy certain 
orthogonality constraints. 
This complicates significantly the construction of such codes.  In
particular, the plain random constructions that work so well in the
classical setting are pointless here. There have been a number of
attempts at overcoming this difficulty and a variety of methods for
constructing quantum LDPC codes have been proposed
\cite{Pos01a,Kit03a,MMM04a,COT05a,COT07a,LG06a,LG08a,HH07a,IofMez07a,Djo08a,SRK08a,Aly07b,Aly08a,HBD08a,TL10a,KHIS11a,CDZ11a,AMT12a,AMT12b}. However,
all of these constructions suffer from disappointingly small
minimum distances, namely whenever they have non-vanishing rate and
parity-check matrices with bounded row-weight, their minimum distance is
either proved to be {\em bounded}, or unknown and with little hope
for unboundedness. 
The point has been made several times that minimum distance
is not everything, because there are complex decoding issues involved,
whose behavior depends only in part on the minimum distance, and also
because a poor asymptotic behavior may be acceptable when one limits
oneself to practical lengths. This is illustrated for instance in our case by the codes constructed in  \cite{KHIS11a,AMT12b} whose 
performance under iterative decoding are outstanding. Nevertheless, 
very poor minimum distances will imply significant error
  floors. We note also that it has recently been proved \cite{KP12b}
  that a sufficiently large growing minimum distance -- for quantum LDPC codes
  -- is enough to
  imply a non-zero decoding threshold, i.e. that the code corrects
  almost all error patterns of weight up to a value linear in the
  block length. Finally, 
the minimum distance has been
the most studied parameter of error-correcting codes and 
given that
asymptotically good (dimension and minimum distance both linear in the
blocklength) quantum LDPC codes are expected to exist, 
it is of great
theoretical interest, and possibly also practical, to devise quantum
LDPC codes with large, growing, minimum distance. This is the problem
that
we address in the present paper, leaving aside decoding issues for
discussion elsewhere.

 Besides the above constructions,
 we must mention the design of quantum LDPC codes based on tessellations
of surfaces \cite{Kit03a,BM06a,BM07a,APS08a,Sar10a}, among which the most prominent example is the toric code of \cite{Kit03a}.
Toric codes have minimum distances which grow like the square root
of the blocklength and parity-check equations of weight 4 but
unfortunately have fixed dimension which is $2$, and hence zero
rate asymptotically.
It turns out that by taking appropriate surfaces of large genus, 
quantum LDPC codes of non vanishing rate can be 
constructed with minimum distance logarithmic in the blocklength, 
this has actually been  achieved in \cite[Th. 12.4]{FML02a}, 
see also \cite{Zem09a}, \cite{Kim07a}.
To the best of our knowledge, this is
until now the
only known family of quantum LDPC codes of non-vanishing rate that yields a
(slowly) growing minimum distance.

We improve here on these surface codes  in several ways,  
by providing a flexible construction of quantum LDPC codes from any pair $(\pc_1,\pc_2)$ of parity-check matrices of 
binary LDPC codes $\Co_1$ and $\Co_2$. Although the constructed
quantum code belongs to the CSS class \cite{CS96a,Ste96b}, there is no
restriction  on $\Co_1$ 
and $\Co_2$. For instance, they do not need to be mutually orthogonal spaces as in 
the CSS construction. In particular we can choose $\Co_1=\Co_2$, in which
case our main result reads:

\begin{theorem}\label{th:main_intro}
  Let $\HH$ be a full-rank $(n-k)\times n$ parity-check matrix of a
  classical LDPC code $\Co$ of parameters $[n,k,d]$. There is a
  construction of a quantum LDPC code with $\HH$ as building block, of
  length $N=n^2+(n-k)^2$, dimension $k^2$, and quantum minimum
  distance $d$. The quantum code has a stabilizer (parity-check) matrix with row
  weights of the form $i+j$, where $i$ and $j$ are respectively row
  and column weights of the original parity-check matrix $\HH$.
\end{theorem}

In particular, any family of classical asymptotically good LDPC codes
of fixed rate yields a family of quantum LDPC codes of fixed rate and
minimum distance proportional to a square root of the block length.

It should also be mentioned that the rate of this construction can be
further improved
while keeping the same minimum distance as has been observed in~\cite{KP12a}.
\section{Basic facts about CSS codes and Tanner graphs}

In this section, we recall a few basic facts about quantum codes and give some 
terminology about LDPC codes.

\paragraph{\bf CSS codes.} The codes constructed in this paper fall into the category of Calderbank-Shor-Steane (CSS) codes \cite{CS96a,Ste96b}
which belong to a more general class of quantum codes called stabilizer codes \cite{Got97a,CRSS98a}.
The first class is described with the help of a pair of mutually orthogonal binary codes, whereas
the second class is given by an additive self-orthogonal code over $GF(4)$ with respect to the trace Hermitian inner product.
Quantum codes on $n$ qubits are  linear subspaces of a Hilbert space of dimension $2^n$ and
do not necessarily have a compact representation in general. The nice feature of stabilizer codes is that they allow 
to define
such a space with the help of a very short representation, which is given here by
 a set of generators of the aforementioned additive code. 
Each generator is viewed as an element of the Pauli group on $n$ qubits and the quantum code
is then nothing but the space stabilized by these Pauli group elements. Moreover, the set of errors that such a quantum code
can correct can also be deduced directly from this discrete representation. For the subclass of CSS 
codes, this representation in terms of additive self-orthogonal codes is equivalent to a representation in terms
of a pair $(\CoX,\CoZ)$ of binary linear codes satisfying the condition
$\CoZ^\perp \subset \CoX$. The {\em quantum minimum distance} of such a CSS code is given by
\begin{eqnarray}
\label{eq:distance}
\dQ & \eqdef &\min \{ \dX, \dZ\}, \;\; \text{where}\\
\dX & \eqdef & \min \{|x|,x \in \CoX \setminus \CoZ^\perp\}, \nonumber\\
\dZ & \eqdef & \min \{|x|,x \in \CoZ \setminus \CoX^\perp\}. \nonumber
\end{eqnarray}
Such a code allows one to protect a subspace of $\kQ$ qubits against errors where
\begin{equation}
\label{eq:def_kQ}
\kQ \eqdef \dim \left(\CoX/\CoZ^\perp\right).
\end{equation}
$\kQ$ is called the {\em quantum dimension} of the CSS code.
Notice that this quantity can be expressed in different ways in order to show its symmetric nature.
\begin{eqnarray}
\dim \left(\CoX/\CoZ^\perp \right) & = & \dim \CoX - \dim (\CoZ^\perp) \label{eq:kQ_difference}\\
& = & \dim (\CoX) + \dim (\CoZ) -n \label{eq:kQ_symmetric}\\
& = & n- \dim (\CoX^\perp) + \dim (\CoZ) -n \nonumber \\
& = & \dim \CoZ - \dim (\CoX^\perp) \nonumber \\
& = & \dim \left(\CoZ/\CoX^\perp \right), \label{eq:kQ_other}
\end{eqnarray}
where $n$ denotes the length of $\CoX$ (or of $\CoZ$).

If $\HH_X$ and $\HH_Z$ are parity-check matrices of the binary codes
$\Co_X$ and $\Co_Z$ respectively, the pair $(\HH_X,\HH_Z)$ is referred
to either as the {\em stabilizer matrix} or as the parity-check matrix
of the quantum code, by analogy with the classical case. Its rows are
also referred to as {\em generators} (of the stabilizer group).

\paragraph{\bf LDPC codes.} 
LDPC (Low Density Parity Check) codes are linear codes which have a sparse parity-check matrix. They can be decoded by using the {\em Tanner graph}
associated to such a parity-check matrix $\HH$. This graph is defined as follows. Assume that $\HH=(h_{ij})_{{\substack{1 \leq i \leq r \\ 1 \leq j \leq n}}}$ is an $r \times n$ matrix (where $n$ is the length of 
the code). The Tanner graph, which is denoted by   $\Tan{V}{C}{E}$ is bipartite and has:\\
\begin{itemize}
\item[(i)]  vertex set $V \cup C$, where
the first set $V$ is in bijection with the indices of the columns of $\HH$, say $V=\{1,\dots,n\}$ and is called the
set of {\em variable nodes}, whereas the second set $C$ is called the set of {\em check nodes} and is in bijection
with the indices of the rows of $\HH$: $C=\{\oplus_1,\dots,\oplus_r\}$. \\
\item[(ii)] edge set $E$; there is an edge between $\oplus_i$ and $j$
if and only if $h_{ij}=1$. 
\end{itemize}
A CSS code defined by a couple of binary code $(\CoX,\CoZ)$ is said to
be a {\em quantum LDPC code} if $\CoX$ and $\CoZ$ are LDPC
codes, i.e. if $\CoX$ and $\CoZ$ have parity-check matrices 
$\HH_X$ and $\HH_Z$ that are both sparse.

\section{The toric code and its generalization}

Our construction borrows both from classical LDPCs and Kitaev's toric
quantum code~\cite{Kit03a}. It is not a coincidence that we obtain
minimum distances that grow like the square root of the blocklength,
similarly to the toric code, but we shall achieve much larger
dimensions that can grow linearly in the blocklength.
To get a clear picture of the construction it is
desirable to take a close look at the toric code and explain how we
shall generalize it.

\begin{figure}
  \centering
  \begin{tikzpicture}[scale=0.9]
    \draw (0,0) -- (5,0);
    \draw (0,1) -- (5,1);
    \draw (0,2) -- (5,2);
    \draw (0,3) -- (5,3);
    \draw (0,4) -- (5,4);
    \draw (0,5) -- (5,5);

    \draw (0,0) -- (0,5);
    \draw (1,0) -- (1,5);
    \draw (2,0) -- (2,5);
    \draw (3,0) -- (3,5);
    \draw (4,0) -- (4,5);
    \draw (5,0) -- (5,5);

    \draw[line width=1mm] (2,0) -- (2,5);
    \draw[line width=1mm] (0,1) -- (5,1);
  \end{tikzpicture}
  \caption{a two-dimensional torus: identify opposing sides of the
    outer square.}
  \label{fig:torus}
\end{figure}
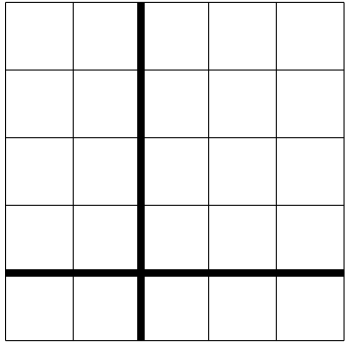

\paragraph{\bf The toric code.}
The toric code is based on the graph $\G$ represented on
Figure~\ref{fig:torus} which is a tiling of the $2$-dimensional torus.
The vertex set of the graph is $\V = \z/m\z\times\z/m\z$ and there is an
edge between every vertex $(x,y)$ and the four vertices $(x\pm 1,y)$,
$(x,y\pm 1)$. In the whole subsection, addition and subtraction are performed modulo $m$. 
Now number the edges from $1$ to $n=2m^2$ so as to
identify the edge set with 
$\{1,2,\dots,n\}$. 
The ambient space $\f_2^n$ is
therefore identified with subsets of edges. The matrix $\HH_X=(h_{ij})$ is the
vertex-edge incident matrix, rows are indexed by vertices of the graph
$\G$, and $h_{ij}=1$ iff vertex $i$ is incident to edge $j$. The
associated code $\Co_X$ is the {\em cycle code} of $\G$,  a cycle being
by definition a set of edges $\Z$ such that every vertex is incident
to an even number of edges of $\Z$. Elements of the row-space
$\Co_X^\perp$ are called {\em cocycles}, rows of $\HH_X$ are called
{\em elementary cocycles}, and the row-space itself $\Co_X^\perp$ is
also known as the {\em cocycle code} of $\G$.

The second matrix $\HH_Z$ of the quantum code is defined as the
face-edge incidence matrix. The faces are defined as the $4$-cycles
$(x,y),(x+1,y),(x+1,y+1),(x,y+1)$.

The rowspace $\Co_Z^{\perp}$ of $\HH_Z$ is therefore a subspace of the
cycle code $\Co_X$, and the quotient $\Co_X/\Co_Z^{\perp}$ is readily seen
to have dimension $2$, coset leaders of the quotient being given by
cycles of the form $(a,0),(a,1),\ldots ,(a,m-1)$ and
$(0,a),(1,a),\ldots , (m-1,a)$, as represented by the thick lines 
on Figure~\ref{fig:torus}. The dimension of the quantum code is
therefore equal to $2$ and the minimum weight of a vector of $\Co_X$ not
in $\Co_Z^{\perp}$ is therefore equal to $m$.

To conclude that the minimum distance of the quantum code is actually
$m$, it remains to determine the minimum weight of a vector of $\Co_Z$
that is not in $\Co_X^{\perp}$, i.e. that is not a cocycle. 
This particular
graph $\G$ has the nice property of being a tiling of a surface (the
torus). This embedding into a surface allows one to define its {\em dual graph}. The (Poincar\'e) dual graph
$\G'$ has vertex set equal to the faces of $\G$, and there is an edge
between two vertices of $\G'$ if the corresponding faces of $\G$ have
a common edge in $\G$. Furthermore the dual graph $\G'$ of $\G$ is
isomorphic to $\G$ itself, and given that the edges of $\G$ define the
edges of $\G'$, the ambient space $\f_2^n$ can be identified with the
edge set of the dual graph $\G'$. With this identification, the
elementary cocycles of $\G$ become the faces of $\G'$ and the faces
of $\G$ become the elementary cocycles of $\G'$. Hence the minimum
weight of a vector of $\Co_Z$
that is not in $\Co_X^{\perp}$ is exactly the same as the minimum weight of a vector of $\Co_X$ not
in $\Co_Z^{\perp}$ and the minimum distance of the quantum code is
exactly $m$.

This duality argument is quite powerful because it ensures that
whatever we prove on the weight of codewords of $\Co_X$ not
in $\Co_Z^{\perp}$ is also valid for the weight of codewords of
$\Co_Z$ not in $\Co_X^{\perp}$: for this reason a number
of quantum codes that arise by replacing the graph $\G$ by different
tilings of different surfaces have been investigated (surface codes).
Here we shall consider a different generalization that does not
destroy graph duality but generalizes~it.

Our first remark is that the graph $\G$ is a product graph: it is the
product of two graphs each equal to an elementary cycle of length $m$.
The (Cartesian) product of two graphs is namely defined as follows.
\begin{definition}[graph product]\label{def:graphproduct}
The product $\gprod{\G_1}{\G_2}$
of two graphs $\G_1$ and $\G_2$ has vertex set made up of
couples $(x,y)$, where $x$ is a vertex of $\G_1$ and $y$ of $\G_2$.
The edges of the product graph connect two vertices $(x,y)$ and
$(x',y')$ if either $x=x'$ and $\{y,y'\}$ is an edge of $\G_2$ or
$y=y'$ and $\{x,x'\}$ is an edge of $\G_1$.
\end{definition}
 Note that any two edges
$\{a,b\}$ and $\{x,y\}$ of $\G_1$ and $\G_2$ define the $4$-cycle of
$\gprod{\G_1}{\G_2}$~:

\begin{center}
  \begin{tikzpicture}[scale=1.6]
    \node (ax) at (0,0) {$(a,x)$};
    \node (ay) at (0,1) {$(a,y)$};
    \node (bx) at (1,0) {$(b,x)$};
    \node (by) at (1,1) {$(b,y)$};
   \draw (ax) -- (ay) -- (by) -- (bx) -- (ax);
    \node (*) at (2,0.5) {({\Large${\mathbf \star}$})};
  \end{tikzpicture}
\end{center}

Now, we are tempted to define a quantum code by, as before, declaring
$\HH_X$ to be the vertex-edge incident matrix of a product graph
$\G=\gprod{\G_1}{\G_2}$ of two arbitrary graphs, and by declaring $\HH_Z$ to
be the matrix whose rows are the characteristic vectors of all faces, i.e.
the $4$-cycles of the form ({\Large${\mathbf \star}$}). This is a quantum code
which generalizes the toric code, since the latter corresponds to the
case when $\G_1$ and $\G_2$ are two cycles of length $m$.
This construction loses graph duality however, and our objective was
to preserve it. In particular, it is not clear at all why the argument used for obtaining the value of $\dX$ can also be used for
deriving $\dZ$. But a closer look shows us that graph duality has not
completely gone: the dual has simply become a {\em hypergraph}, whose vertex
set is  the set of faces of $\G$ and where the hyperedges are the
subsets of those faces of $\G$ that meet in a common edge of $\G$.
This observation shows us that we really should consider products of
hypergraphs rather than graph products to start~with. 
This is the approach which was followed in a preliminary version of
this work~\cite{TZ09a} 
where the construction is described in terms of products of hypergraphs.
This way of viewing the toric code highlights the connections with algebraic topology and is also
a natural generalization of the quantum code construction known under the name of surface codes.

However, we shall slightly change our point of view and detail our
construction in a way that stays with the more familiar notion of graph
product. This is the approach which we follow in the next paragraphs.
The hypergraph connection will become apparent in
Sections~\ref{sec:hypergraph} and~\ref{sec:transpose}.

\paragraph{\bf Another way of viewing the toric code in terms of  a graph product construction.}

\begin{figure}[h!]
\centering
\includegraphics[height=6cm]{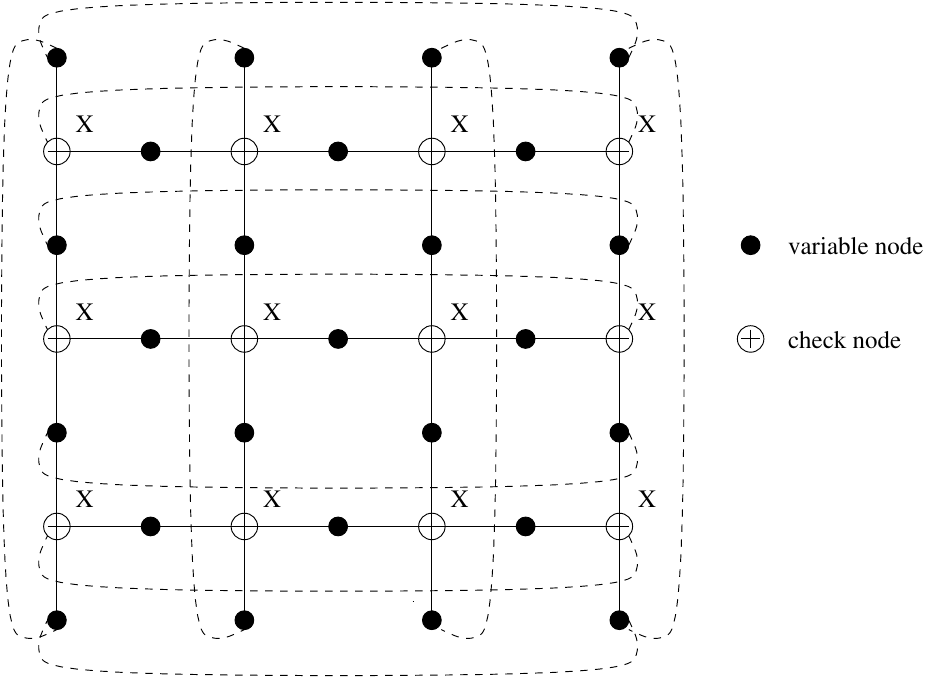}
\caption{The Tanner graph of $\CoX$ for $m=3$: the dashed lines indicate the pair of vertices which are identified. }
  \label{fig:TannerX}
\end{figure}

There arises another interpretation of the toric code as a product of
two cycles, of length double the previous length, by considering
simultaneously the Tanner graphs of $\CoX$ and $\CoZ$ and by putting the qubits on the vertices instead of the edges. Indeed, let us consider for instance the Tanner graph corresponding 
to $\HH_X$ for $m=3$ which is depicted in Figure \ref{fig:TannerX}.

\begin{figure}[h!]
\centering
\includegraphics[height=7cm]{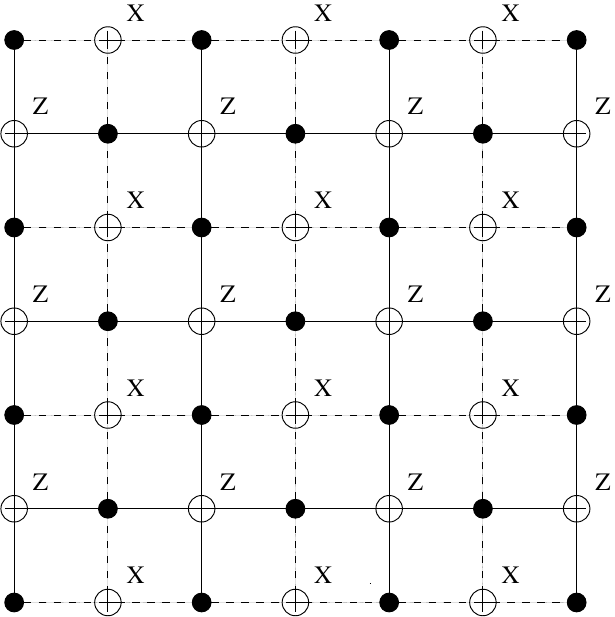}
\caption{The union of the Tanner graph of $\CoX$ and $\CoZ$ for $m=3$: identify the opposing vertices of the
    outer square, the dashed lines indicate here the edges of the
    Tanner graph of $\CoZ$, the solid lines the edges of the
    Tanner graph of $\CoX$.  The 
    check nodes marked with an ``$X$'', respectively with a ``$Z$''  belong to the Tanner graph of $\CoX$, 
    respectively $\CoZ$.}
  \label{fig:TannerXZ}
\end{figure}

It is insightful to consider the union of both Tanner graphs of $\CoX$ and $\CoZ$ (see Figure \ref{fig:TannerXZ}).

\begin{figure}[h!]
\centering
\includegraphics[height=2.5cm]{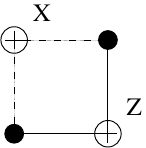}
\caption{Two check nodes of different types sharing two different variable nodes.}
  \label{fig:square}
\end{figure}

The Tanner graph of $\CoX$ does not have a product graph structure, whereas
the union of both Tanner graphs has now the structure of the product of two cycles
of length $6$ in the example and $2m$ in general. Notice that such a cycle can be viewed as the Tanner graph 
of the repetition code of length $m$. Moreover, it is clear from this picture why the rows of $\HH_X$ and
$\HH_Z$ are orthogonal : two parity-check nodes of type $X$ and $Z$ respectively which are adjacent to a same variable
node are also adjacent to a second variable node as shown in Figure \ref{fig:square}. This comes from the very definition
of a product graph as explained before. 

This discussion strongly suggests to consider the following generalization of the toric code.
\begin{definition}{\bf (CSS code $\QC{\G_1}{\G_2}$ 
associated to a graph product)}\label{def:Q}

Let  $\G_1=\Tan{V_1}{C_1}{E_1}$ and $\G_2=\Tan{V_2}{C_2}{E_2}$ be two
Tanner graphs.
Let 
\begin{align*}
  V &\eqdef V_1 \times V_2 \cup C_1 \times C_2\\
  C &\eqdef C_1 \times V_2 \cup V_1 \times C_2,
\end{align*}
so that the product
graph $\G_1\times\G_2$ is a bipartite graph with vertex set $V\cup C$.
Let $\GXd{\G_1}{\G_2}$ be the Tanner graph defined as the subgraph of
$\gprod{\G_1}{\G_2}$ with set of variable nodes  $V$ 
and set of check nodes $C_1 \times V_2$. Similarly, define the Tanner
graph $\GZd{\G_1}{\G_2}$ as the subgraph of $\gprod{\G_1}{\G_2}$ with 
set of variable nodes $V$ and set of check nodes $V_1 \times C_2$: the
union of the two edge-sets of $\GXd{\G_1}{\G_2}$ and
$\GZd{\G_1}{\G_2}$ make up therefore the edge set of
$\gprod{\G_1}{\G_2}$.
Finally, define the two classical codes
$\Co_X=\Co_X(\gprod{\G_1}{\G_2})$ and 
$\Co_Z=\Co_Z(\gprod{\G_1}{\G_2})$ as the
codes associated to the Tanner graphs $\GXd{\G_1}{\G_2}$ and
$\GZd{\G_1}{\G_2}$ respectively.

The couple $(\Co_X,\Co_Z)$ defines a CSS code that we denote by $\QC{\G_1}{\G_2}$.
\end{definition}

\begin{figure}[h!]
\centering
\begin{tikzpicture}[scale=0.9]
  \draw (0,0) ellipse (0.7cm and 2cm);
  \draw (2,0) ellipse (0.6cm and 1.5cm);
  \draw (0,5) ellipse (0.7cm and 2cm);
  \draw (2,5) ellipse (0.6cm and 1.5cm);
  \begin{scope}
    \tikzstyle{every node}=[circle, draw, fill=black!100,
                        inner sep=0pt, minimum width=4pt]
   \node (a) at (0,1.6) {};
   \node (b) at (0,1.2) {};
   \node (c) at (0,0.8) [label=below:$v_1$] {};
   \node (d) at (0,-1.6) {};

   \node (a') at (0,6.6) {};
   \node (b') at (0,6.2) {};
   \node (c') at (0,5.8) [label=below:$v_2$] {};
   \node (d') at (0,3.6) {};
  \end{scope}
  \begin{scope}
   \tikzstyle{every node}=[circle, draw,
                        inner sep=0pt, minimum width=4pt]
    \node (e) at (2,1.2) {};
    \node (f) at (2,0.8) {};
    \node (g) at (2,0.4) [label=below right:$c_1$] {};
    \node (h) at (2,-1.2) {};

    \node (e') at (2,6.2) {};
    \node (f') at (2,5.8) {};
    \node (g') at (2,5.4) [label=below right:$c_2$] {};
    \node (h') at (2,4.2) {};
  \end{scope}

   \node (V1) at (-0.7,2.2) {{\LARGE $V_1$}};
   \node (V2) at (-0.7,7.2) {{\LARGE $V_2$}};
   \node (C1) at (1.3,1.7) {{\LARGE $C_1$}};
   \node (C2) at (1.3,6.7) {{\LARGE $C_2$}};

\draw [decorate,decoration={brace,amplitude=10pt,mirror,raise=4pt},yshift=0pt]
(2.8,-1.5) -- (2.8,6.5) node (x) [black,midway,xshift=7mm] {};
\draw[->,thick] (x) -- node[above] {{\LARGE $\times$}} (5,2.5);

   \draw (7,2.5) ellipse (1.5cm and 4cm);
   \draw (11,2.5) ellipse (1.2cm and 3.5cm);
   \draw (7,4.5) ellipse (0.9cm and 1.7cm);
   \draw (7,0.4) ellipse (0.9cm and 1.5cm);

   \draw (11,4.1) ellipse (0.9cm and 1.2cm);
   \draw (11,1) ellipse (0.9cm and 1.2cm);
   \node (VV) at (7,3.9) {{\large $V_1\times V_2$}};
   \node (CC) at (7,0) {{\large $C_1\times C_2$}};
   \node (CV) at (11,3.6) {{\large $C_1\times V_2$}};
   \node (VC) at (11,0.5) {{\large $V_1\times C_2$}};
   \begin{scope}
   \tikzstyle{every node}=[circle, draw, fill=black!100, inner sep=0pt, minimum width=4pt]
   \node (vv) at (7,5) {};  
   \node (cc) at (7.1,0.5) {};
   \end{scope}
   \begin{scope}
    \tikzstyle{every node}=[circle, draw,
                        inner sep=0pt, minimum width=4pt]
   \node (cv) at (10.8,4.4) {};
   \node (vc) at (10.9,1) {};
   \end{scope}
   \node (vvlabel) at (7,5.4) {$(v_1,v_2)$};
   \node (cclabel) at (6.9,1.2) {$(c_1,c_2)$};
   \node (cvlabel) at (11,4.8) {$(c_1,v_2)$};
   \node (vclabel) at (11.1,1.6) {$(v_1,c_2)$};

   \draw (a) -- (e) (a) -- (f) (c) -- (e) (d) -- (h) ;
   \draw (a') -- (e') (a') -- (f') (c') -- (e') (d') -- (h');

   \draw (vv) -- node[above,sloped] {${\mathbf X}$} (cv) (cc) --
   node[above,sloped] {${\mathbf X}$} (cv);
   \draw[style=dashed] (vv) -- node[above,sloped] {${\mathbf Z}$} (vc) (cc) -- node[above,sloped] {${\mathbf Z}$} (vc);
\end{tikzpicture}
\caption{The Tanner graphs of $\CoXd{\G_1}{\G_2}$ and $\CoZd{\G_1}{\G_2}$ are given by the edges of $\G_1 \times \G_2$ which 
join $V$ to $C_1 \times V_2$ for $\Co_X$ (the corresponding edges are denoted by solid lines) and which join
$V$ to $V_1 \times C_2$ for $\Co_Z$ (the corresponding edges are indicated by dashed lines). }
  \label{fig:CSS_product}
\end{figure}
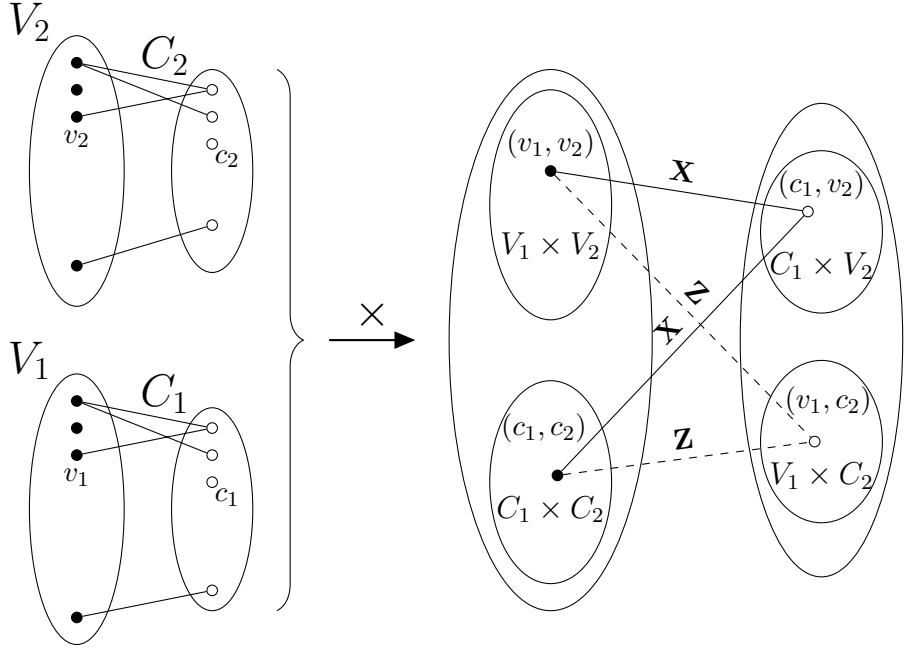

The construction is summarized on Figure \ref{fig:CSS_product}. 

It is readily checked that if $\G_1=\G_2$ are two cycles of even
length $2m$, so that $\G_1$ and $\G_2$ are two isomorphic bipartite
graphs with $|V_1|=|V_2|=|C_1|=|C_2|=m$, then the two Tanner graphs 
$\GXd{\G_1}{\G_2}$ and $\GZd{\G_1}{\G_2}$ are the familiar Tanner graphs
of $\Co_X$ and $\Co_Z$ depicted on Figure~\ref{fig:TannerX}.

It will be explained shortly in the next section why the 
$4$-cycle joining $(v_1,v_2),(c_1,v_2),(c_1,c_2)$ and $(v_1,c_2)$ 
ensures that the couple $(\Co_X,\Co_Z)$ always defines indeed a CSS code.
It will also turn out that the minimum distance of the quantum 
code is related to the minimum distance of the classical binary codes 
with Tanner graphs $\G_1$ and $\G_2$.

\section{Dimension of the CSS code $\QC{\G_1}{\G_2}$ and relationship with product codes}
\label{sec:dim}

\subsection{Validity of the construction of $\QC{\G_1}{\G_2}$.} 
We prove here that $\QC{\G_1}{\G_2}$ is indeed a CSS code. This follows at once from
\begin{proposition}
Let  $\G_1=\Tan{V_1}{C_1}{E_1}$ and $\G_2=\Tan{V_2}{C_2}{E_2}$ be two
Tanner graphs. We have:
$$
\CoXd{\G_1}{ \G_2}^\perp \subset \CoZd{\G_1}{\G_2}
$$
\end{proposition}

\begin{proof}{}
Let $\HH_X$ and $\HH_Z$ be the parity-check matrices associated to the
Tanner graphs $\GXd{\G_1}{\G_2}$ and $\GZd{\G_1}{\G_2}$ respectively.
For $c_1\in C_1$ and $v_2\in V_2$,
denote by $\h_X(c_1,v_2)$ the row of $\HH_X$ corresponding to the check
node $(c_1,v_2)$ of $\GXd{\G_1}{\G_2}$. It will be convenient to
view vectors as sets of (variable) nodes of $\gprod{\G_1}{\G_2}$ by
identifying them with their supports, so that $\h_X(c_1,v_2)$ is the
set of neighbors of $(c_1,v_2)$ in $\gprod{\G_1}{\G_2}$. Similarly,
row $\h_Z(v_1,c_2)$ of $\HH_Z$ should be thought of the set of
neighbors of $(v_1,c_2)$ in the graph $\gprod{\G_1}{\G_2}$, for some 
$v_1\in V_1$ and $c_2\in C_2$.

To prove that $\CoXd{\G_1}{ \G_2}^\perp \subset \CoZd{\G_1}{\G_2}$ it
is sufficient to prove that for any $v_i\in V_i$, $c_i\in C_i$,
$i\in\{1,2\}$,
row $\h_X(c_1,v_2)$ of $\HH_X$ is orthogonal to row $\h_Z(v_1,c_2)$ of $\HH_Z$.
This is achieved as follows. We first notice that the inner product 
between 
$\h_X(c_1,v_2)$ and $\h_Z(v_1,c_2)$ can be expressed as 
$$\langle \h_X(c_1,v_2),\h_Z(v_1,c_2)\rangle = \# S \pmod{2},$$
where $S$ is 
the set of elements of $V_1 \times V_2 \cup C_1 \times C_2$ which are
adjacent to both $(c_1,v_2)$ and $(v_1,c_2)$ in $\G_1 \times \G_2$.
Now the set $S$ is clearly empty if either $v_1$ is not adjacent to $c_1$ in $\G_1$ or if $c_2$ is not adjacent to $v_2$ in $\G_2$.
When $v_i$ is adjacent to $c_i$ in $\G_i$ for $i=1,2$, then there are exactly two vertices in $V_1 \times V_2 \cup C_1 \times C_2$
which are both adjacent to $(c_1,v_2)$ and $(v_1,c_2)$, namely
$(c_1,c_2)$ and $(v_1,v_2)$. This implies in both cases that 
$\h_X(c_1,v_2)$ and $\h_Z(v_1,c_2)$ are orthogonal.
\end{proof}

\subsection{Degree structure of the Tanner graphs $\GXd{\G_1}{\G_2}$ 
and $\GZd{\G_1}{\G_2}$ of 
$\CoXd{\G_1}{ \G_2}$ and $\CoZd{\G_1}{ \G_2}$.}  The Tanner graphs of all the constructions of quantum LDPC codes which have been proposed so far
have in general a very specific degree structure. Unfortunately it is well known that in order to obtain classical LDPC codes
which allow to operate successfully at rates very close to capacity, the degree structure has to be optimized very carefully. This kind of approach can not 
be carried out for any of  the constructions which have been mentioned in the introduction but one (namely the 
 the construction based on LDGM codes \cite{LG06a,LG08a}). However in this case, the minimum distance is constant and there is no threshold
error probability for decoding. The construction which is proposed here allows in principle this kind of optimization up to some extent as we will see now.

\begin{proposition}
\label{pr:degree}
Let $(\lambda_i(j))_j,(\rho_i(j))_j$ be respectively the right and left degree distributions of the Tanner graph $\G_i= \Tan{V_i }{C_i}{E_i}$ for $i \in \{1,2\}$
(that is $\lambda_i(j)$ is the fraction of vertices of $V_i$ of degree $j$, whereas $\rho_i(j)$ is the fraction of vertices of $C_i$ of degree $j$).
We let $\lambda_X(j)$, respectively $\rho_X(j)$, be the fraction of vertices of  $V_1 \times V_2 \cup C_1 \times C_2$, respectively of
$C_1 \times V_2$ of degree $j$ in
$\GXd{\G_1}{\G_2}$. 
$\lambda_Z(j)$, respectively $\rho_Z(j)$, are defined similarly as 
the fraction of vertices of  $V_1 \times V_2 \cup C_1 \times C_2$, respectively of
$V_1 \times C_2$ of degree $j$ in
$\GZd{\G_1}{\G_2}$. 
For $i \in \{1,2\}$  we define the following average left and right degrees
\begin{eqnarray}
\overline{\lambda_i} & \eqdef & \sum_j \lambda_i(j) j\\
\overline{\rho_i} & \eqdef & \sum_j \rho_i(j) j
\end{eqnarray}
We have
\begin{eqnarray}
\lambda_X(j) & = & \frac{ \lambda_1(j) + \frac{\up{\lambda_1}\up{\lambda_2}}{\up{\rho_1}\up{\rho_2}}\rho_2(j)}
{\frac{\up{\lambda_1}\up{\lambda_2}}{\up{\rho_1}\up{\rho_2}} + 1} \\
\lambda_Z(j) & = & \frac{ \lambda_2(j) + \frac{\up{\lambda_1}\up{\lambda_2}}{\up{\rho_1}\up{\rho_2}}\rho_1(j)}
{\frac{\up{\lambda_1}\up{\lambda_2}}{\up{\rho_1}\up{\rho_2}} + 1} \\
\rho_X(k) & = & \sum_{i,j:i+j=k} \rho_1(i) \lambda_2(j) \\
\rho_Z(k) & = & \sum_{i,j:i+j=k} \lambda_1(i) \rho_2(j)
\end{eqnarray}
\end{proposition}

\begin{proof}{}
The number of vertices of degree $j$ in $V_1 \times V_2$ of the graph $\GXd{\G_1}{\G_2}$ is 
equal to $|V_1| |V_2| \lambda_1(j)$, whereas the number of vertices of degree $j$ in $C_1 \times C_2$ of degree 
$j$ is equal to $|C_1| |C_2| \rho_2(j)$. This implies that
\begin{eqnarray*}
\lambda_X(j) & =& \frac{|V_1| |V_2| \lambda_1(j)+ |C_1| |C_2| \rho_2(j)}{|V_1| |V_2|+ |C_1| |C_2|} \\
& = & \frac{ \lambda_1(j) + \frac{\up{\lambda_1}\up{\lambda_2}}{\up{\rho_1}\up{\rho_2}}\rho_2(j)}
{\frac{\up{\lambda_1}\up{\lambda_2}}{\up{\rho_1}\up{\rho_2}} + 1}
\end{eqnarray*}
since $\frac{|C_i]}{|V_i|} = \frac{\up{\lambda_i}}{\up{\rho_i}}$.
The number of vertices of degree $k$ in $C_1 \times V_2$ of the  graph $\GXd{\G_1}{\G_2}$
is equal to the number of vertices $(c_1,v_2) \in C_1 \times V_2$ such that
the degree of $c_1$ in $\G_1$ plus the degree of $v_2$ in $\G_2$ is equal to $k$. This number is therefore equal to
$\sum_{i,j:i+j=k} |C_1||V_2| \rho_1(i) \lambda_2(j).$ This yields immediately that
$$
\rho_X(k)  =  \sum_{i,j:i+j=k} \rho_1(i) \lambda_2(j).
$$
The formulas for $\lambda_Z(j)$ and $\rho_Z(k)$ are obtained in a similar fashion.
\end{proof}

This result implies that there is a large
degree of freedom for choosing the degree distributions of the Tanner graphs $\GXd{\G_1}{\G_2}$ and
$\GZd{\G_1}{\G_2}$. This can potentially be used to devise quantum LDPC codes with very good iterative decoding performances.

\subsection{The hypergraph connection, product codes}\label{sec:hypergraph}
Our objective is to derive a formula for the dimension of the quantum
code $\QC{\G_1}{\G_2}$. For this purpose we shall relate the Tanner graphs
$\GXd{\G_1}{\G_2}$ and $\GZd{\G_1}{\G_2}$ to Tanner graphs of product
codes.

We first need to define another product notion $\G_1\otimes\G_2$ for
two Tanner graphs $\G_1$ and $\G_2$ that will make $\G_1\otimes\G_2$ the
Tanner graph of the product code $\Co_1\otimes \Co_2$ when $\Co_1$ and
$\Co_2$ are the codes associated to the Tanner graphs $\G_1$ and
$\G_2$. 
This product is the natural extension of
Definition~\ref{def:graphproduct} when $\G_1$ and $\G_2$ are viewed as
{\em hypergraphs}.

Recall that a {\em hypergraphs} $\H=(V,\E)$ is simply a set $V$ together
with a collection $\E$ of subsets called {\em hyperedges} or simply
{\em edges}. A hypergraph is a graph when every edge has cardinality
$2$. Consider the following definition:

\begin{definition}\label{def:hypergraphproduct}
  Let $\H_1=(V_1,\E_1)$ and $\H_2=(V_2,\E_2)$ be two hypergraphs. The
  {\em product hypergraph} $\H_1\times \H_2$ is defined as the hypergraph
  $\H=(V,\E)$ such that $V=V_1\times V_2$ and $\E$ is the collection
  of subsets of $V$ 
  \begin{itemize}
  \item either of the form $\{v_1\}\times e_2$ with $v_1\in V_1$ and
    $e_2\in\E_2$,
  \item or of the form $e_1\times \{v_2\}$ with $e_1\in\E_1$ and $v_2\in V_2$.
  \end{itemize}
\end{definition}

It should be clear that when $\H_1$ and $\H_2$ are graphs then
$\H_1\times \H_2$ reduces to the graph product of
Definition~\ref{def:graphproduct}.

Now hypergraphs are often described by their associated bipartite
graph: for a hypergraph $\H=(V,\E)$ this bipartite graph has vertex
set $V\cup\E$ and we put an edge between $v\in V$ and $e\in \E$
whenever $v\in e$ in $\H$. Conversely, any bipartite graph, in other
words any Tanner graph $\Tan{V}{C}{E}$, generates a hypergraph on
vertex set $V$ by declaring that the hyperedges are the
neighborhoods of the check vertices $c$, when $c$ ranges over $C$.

Tanner graphs are of course simply bipartite graphs. The terminology
``Tanner'' serves the sole purpose of reminding us that we are
interested in the associated error-correcting code.
Translated into Tanner graph terminology,
Definition~\ref{def:hypergraphproduct} becomes:

\begin{definition}
  Let $\G_1=\Tan{V_1}{C_1}{E_1}$ and $\G_2=\Tan{V_2}{C_2}{E_2}$ be two
  Tanner graphs. The {\em hypergraph product} $\G_1\otimes\G_2$ of
  $\G_1$ and $\G_2$ is defined as the induced subgraph of $\G_1\times
  \G_2$ with variable node  set $V_1 \times V_2$ and 
check node set $C_1 \times V_2 \cup V_1 \times C_2$.
\end{definition}

Now the hypergraph product of Tanner graphs is directly related to the
standard product construction for codes. We recall the definition:

\begin{definition}[Product code]
Let $\Co_1$ and $\Co_2$ be two binary codes of length $n_1$ and $n_2$ respectively. The product code $\Co_1 \otimes \Co_2$ of 
$\Co_1$ and $\Co_2$ is the binary code of length $n_1 \times n_2$ whose codewords may be viewed as binary matrices of size $n_1 \times n_2$
and such that a matrix belongs to $\Co_1 \otimes \Co_2$ if and only if all its columns belong to $\Co_1$ and all its rows to $\Co_2$.
\end{definition}

It is well known (see \cite[Ch. 18. §2]{MacSlo} for instance) that the dimension of the product code is given by
\begin{proposition}
\label{pr:dimension_product}
Let $\Co_1$ and $\Co_2$ be two binary linear  codes.
$$
\dim(\Co_1 \otimes \Co_2) = \dim \Co_1 \dim \Co_2.
$$
\end{proposition}

The following proposition is straightforward:

\begin{proposition}
For $i \in \{1,2\}$, let
$\G_i=\Tan{V_i}{C_i}{E_i}$ be a Tanner graph of the binary linear code $\Co_i$.
A Tanner graph for $\Co_1 \otimes \Co_2$ is given by $\G_1\otimes\G_2$.
\end{proposition}

Now to describe the Tanner graphs  $\GXd{\G_1}{\G_2}$ and
$\GZd{\G_1}{\G_2}$ in terms of hypergraph products and to derive the
dimensions of the associated codes $\Co_X$ and $\Co_Z$, we need one extra
tool:

\subsection{The Transpose Tanner Graph}\label{sec:transpose}
\begin{definition}[Transpose of a Tanner graph]
The transpose of a Tanner graph $\Tan{V}{C}{E}$ is the Tanner graph $\Tan{C}{V}{E}$.\end{definition}

In other words, transposing a Tanner graph amounts to exchanging the
role of the variable node set with the check node set. For a binary
linear code $\Co$, we shall abuse notation somewhat and 
denote by $\Co^T$ (and call it a transpose
code of $\Co$) the binary code specified by $\G^T$ where $\G$ is a Tanner graph for $\Co$, even if this notion
assumes implicitly some choice for the Tanner graph of $\Co$.
We do this whenever  the choice of the 
underlying Tanner graph is obvious from the context.
Note that the length of $\Co^T$ can be varied by adding or removing
redundant parity-checks to the parity-check matrix/Tanner graph for
$\Co$. 
The transpose code can be viewed as the code associated to the 
linear combinations of the rows of a parity-check matrix of the code which are equal to $0$. This interpretation 
directly leads to the following relationship between the dimension of
a code and its transpose:
\begin{fact}
\label{fa:code_transpose}
Let $\Co$ be a binary code specified by a Tanner graph $\G=\Tan{V}{C}{E}$.
Then the dimension $\dim(\Co)$ of $\Co$ and the dimension $\dim(\Co^T)$ of the code associated to the transpose 
$\G^T$ are related by
\begin{equation}
\label{eq:code_transpose}
\dim(\Co) = |V| - |C| + \dim(\Co^T).
\end{equation}
\end{fact}

We are now ready now to derive the aforementioned relationship between
the Tanner graphs $\GXd{\G_1}{\G_2}$ and $\GZd{\G_1}{\G_2}$ and Tanner graphs of product codes.

\begin{proposition}
\label{pr:product_code}
For $i \in \{1,2\}$, let $\G_i=\Tan{V_i}{C_i}{E_i}$ be the Tanner
graph of a binary code $\Co_i$, then
\begin{align}
  \GXd{\G_1}{\G_2} &= (\G_1^T\otimes \G_2)^T \label{eq:hprod_X}\\
  \GZd{\G_1}{\G_2} &= (\G_1\otimes \G_2^T)^T. \label{eq:hprod_Z}
\end{align}
In other words,
\begin{eqnarray}
\CoXd{\G_1}{\G_2}^T & = & \Co_1^T \otimes \Co_2 \label{eq:prod_X}\\
\CoZd{\G_1}{\G_2}^T & = & \Co_1 \otimes \Co_2^T.   \label{eq:prod_Z}
\end{eqnarray}
\end{proposition} 

\begin{proof}{}
Remember that $\GXd{\G_1}{\G_2}$ is the subgraph of $\G_1 \times \G_2$
induced by vertex node set 
$V_1 \times V_2 \cup C_1 \times C_2$ and check node set $C_1 \times
V_2$. This gives \eqref{eq:hprod_X} by definition of the hypergraph
product: \eqref{eq:hprod_Z} is obtained
analogously. Equalities \eqref{eq:prod_X} and \eqref{eq:prod_Z} follow directly.
\end{proof}

From this last proposition, Proposition \ref{pr:dimension_product} and Fact \ref{fa:code_transpose}, we obtain immediately that
the dimension of the quantum code $\QC{\G_1}{\G_2}$ is given by
\begin{proposition}
\label{pr:dimension} For $i \in \{1,2\}$, let $\G_i$ be the Tanner graph of a binary code $\Co_i$ and
let $k_i = \dim(\Co_i), k_i^T = \dim(\Co_i^T)$. The quantum dimension $\kQ$ of $\QC{\G_1}{\G_2}$ is given by
$$\kQ = k_1 k_2 + k_1^Tk_2^T.$$
\end{proposition}

\begin{proof}{}
Let $r_i$ be the number of check nodes of $\G_i$ and 
let $n_i$ be the length of $\Co_i$.
By using first Fact \ref{fa:code_transpose}, then Proposition \ref{pr:product_code} and finally Proposition \ref{pr:dimension_product},  we obtain
\begin{eqnarray*}
\dim(\CoXd{\G_1}{\G_2}) & = & n_1 n_2 + r_1 r_2 - r_1 n_2 + \dim\left( \CoXd{\G_1}{\G_2}^T \right)\\
& = & n_1 n_2 + r_1 r_2 - r_1 n_2 + \dim\left(\Co_1^T \otimes \Co_2 \right) \\
& = & n_1 n_2 + r_1 r_2 - r_1 n_2 + \dim(\Co_1^T) \dim(\Co_2) \\
& = & n_1 n_2 + r_1 r_2 - r_1 n_2 + k_1^T k_2
\end{eqnarray*}
We derive in a similar way that
$$
\dim(\CoZd{\G_1}{\G_2}) = 
n_1 n_2 + r_1 r_2 - r_2 n_1 + k_2^T k_1
$$
By using Formula \eqref{eq:kQ_symmetric} for the quantum dimension we obtain
\begin{eqnarray*}
\kQ  
&= & \dim(\CoXd{\G_1}{\G_2}) - n_1 n_2 - r_1 r_2 + \dim(\CoZd{\G_1}{\G_2})\\
& = & n_1 n_2 + r_1 r_2 - r_1 n_2 + k_1^T k_2 - n_1 n_2 - r_1 r_2 \\
& & + n_1 n_2 + r_1 r_2 - r_2 n_1 + k_2^T k_1\\
& = & (n_1-r_1)(n_2-r_2) + k_1^T k_2 + k_2^T k_1\\
& = & (k_1-k_1^T)(k_2-k_2^T)+k_1^T k_2 + k_2^T k_1\\
& = & k_1 k_2 + k_1^Tk_2^T.
\end{eqnarray*}
\end{proof}

\paragraph{Comment.} 
If we go back to our initial definition of
Kitaev's toric code, with $\Co_X$ being defined as the cycle code of
the (graph) product of a cycle with itself, and if we wish to express the
Tanner graph of $\Co_X$ as a function of the Tanner graph $\G$ of the original
cycle, then the more natural expression for $\Co_X$ will be
that it has the Tanner graph
  $$(\G\otimes\G)^T$$
and then $\Co_Z$ will come out as:
  $$(\G^T\otimes\G^T)^T.$$
However, since $\G^T$ is isomorphic to $\G$ (a cycle of even length),
the more twisted Tanner graph expressions in \eqref{eq:hprod_X} and
\eqref{eq:hprod_Z} generalize the toric code just as
well. Definition~\ref{def:Q}, equivalently
Proposition~\ref{pr:product_code},
has the following two advantages:
\begin{itemize}
\item the quantum code $\QC{\G}{\G}$ always has
positive dimension for $\G$ the Tanner graph of any non-trivial code,
by Proposition~\ref{pr:dimension},
\item $\Co_X(\G\otimes\G)$ and $\Co_Z(\G\otimes\G)$ are clearly isomorphic.
\end{itemize}

\section{Minimum distance}
In this section, we show that the minimum distance of the quantum code $\QC{\G_1}{\G_2}$ has a very simple expression,
when we adopt the convention that the minimum distance of a code
reduced to the all-zero codeword is $\infty$
\begin{theorem}
\label{th:main}
For $i \in \{1,2\}$, let $d_i$ be the minimum distance of a code
with Tanner graph 
$\G_i$ and let $d_i^T$ denote the minimum distance of the code specified by the transpose Tanner graph
$\G_i^T$. The minimum distance $\dQ$ of the quantum code $\QC{\G_1}{\G_2}$ 
satisfies
$$
\dQ \geq  \min(d_1,d_2,d_1^T,d_2^T).
$$
and is given by
$$
\dQ = \min(d_1,d_2,d_1^T,d_2^T).
$$ 
in the following cases
\begin{itemize}
\item $d_i = \min(d_1,d_2,d_1^T,d_2^T)$ for some $i \in \{1,2\}$ and
$d_{3-i} \neq \infty$,
\item or $d_i^T = \min(d_1,d_2,d_1^T,d_2^T)$ for some $i \in \{1,2\}$ and
$d_{3-i}^T \neq \infty$.
\end{itemize}
\end{theorem}
In other words, the minimum distance of the quantum code is  governed by the minimum distance
of the underlying binary codes $\Co_i$ with Tanner graphs $\G_1$ and $\G_2$ when $\G_1^T$ and $\G_2^T$ 
are Tanner graphs of trivial codes. This also implies that by choosing the binary linear codes
$\Co_i$ with linear minimum distance we will be able to construct families of quantum codes
with minimum distance behaving like the square root of the blocklength (and with a dimension which is
linear in the blocklength when the rates of $\Co_i$ are chosen appropriately). Moreover, by choosing $\G_1$ and $\G_2$ 
to be sparse it turns out by Proposition \ref{pr:degree} that $\GXd{\G_1}{\G_2}$ and $\GZd{\G_1}{\G_2}$ are also sparse.
In other words we obtain in this way quantum LDPC codes with a minimum distance which can be of the form $\Omega(\sqrt{N})$
where $N$ is the blocklength of the quantum code. We will prove this theorem by first proving that the
righthand-side is a lower bound on the minimum distance, then we will prove that this lower bound is attained by exhibiting
suitable codewords of the quantum code.

\subsection{A lower bound on the minimum distance}

A common strategy for obtaining a lower bound on the minimum distance
of a quantum CSS code is to simply look for lower bounds on the
minimum distances of the classical codes $\CoX$ and $\CoZ$. However
this approach will necessarily fail here because of the LDPC nature of
the CSS code. Recall that since $\CoZ^\perp\subset \CoX$ and
$\CoX^\perp\subset \CoZ$, any lower bound on the minimum weight of 
$\CoX$ and $\CoZ$ will not exceed the minimum weight of the
parity-check matrices $\HH_X$ and $\HH_Z$ of $\CoX$ and $\CoZ$, which
is a constant since $\HH_X$ and $\HH_Z$ are precisely constructed to
have small row weights.

To obtain any interesting lower bound on the quantum minimum distance,
we must therefore use its full definition, namely that the minimum
distance is the smallest weight of a codeword of $\CoX$ {\em not in
  $\CoZ^\perp$} or of a codeword of $\CoZ$ {\em not in $\CoX^\perp$}.
Our strategy will therefore be to consider a non-zero element
of $\CoXd{\G_1}{\G_2}$ and prove that when its weight is too small, then it has to belong to $\CoZd{\G_1}{\G_2}^\perp$.

By using the same notation as for Theorem \ref{th:main} we will prove in this way that
\begin{lemma}
\label{lem:lower_bound}
$$\dQ \geq  \min(d_1,d_2,d_1^T,d_2^T).$$
\end{lemma}

\begin{proof}{}
We first prove that any  element $\x$ of $\CoXd{\G_1}{\G_2}$ which is of weight less than
$\min(d_1,d_2^T)$ belongs to $\CoZd{\G_1}{\G_2}^\perp$. We denote by $\supp(\x)$ the support of $\x$
which is a
 subset of $V_1 \times V_2 \cup C_1 \times C_2$. Let $V'_1 \eqdef  \{v' \in V_1 : \exists v \in V_2, (v',v) \in \supp(\x)\}$ and
$C'_2 \eqdef  \{c' \in C_2 : \exists c \in C_1, (c,c') \in \supp(\x)\}$. 
Let $\G'_1$ be the subgraph of $\G_1$ induced by $V'_1 \cup C_1$ and let 
$\G'_2$ be the subgraph of $\G_2$ induced by $V_2 \cup C'_2$. 

Let $\Co_i$ (respectively $\Co'_i$) be the binary code defined by the
Tanner graph $\G_i$ (respectively $\G'_i$), and let
${\Co'_i}^T$ be the binary code described by ${\G'_i}^T$. 
Since a codeword of $\Co'_1$ can be viewed as a codeword of $\Co_1$ by
extending it with zeros on the positions of $V_1 \setminus V'_1$ and since $|V'_1| < d_1$ we necessarily have
$\dim(\Co'_1)=0$. A similar reasoning shows that $\dim({\Co'_2}^T)=0$. By using Proposition \ref{pr:dimension} we see that
the dimension of $\QC{\G'_1}{\G'_2}$ is equal to zero. This is equivalent to 
\begin{equation}
\label{eq:equality}
\CoXd{\G'_1}{\G'_2}=\CoZd{\G'_1}{\G'_2}^\perp.
\end{equation}
Notice now that  the restriction $\x'$ of $\x$ to the positions in $V'_1 \times V_2 \cup C_1 \times C'_2$ belongs to $\CoXd{\G'_1}{\G'_2}$.
Therefore it also belongs to $\CoZd{\G'_1}{\G'_2}^\perp$ and therefore $\x$ belongs to $\CoZd{\G_1}{\G_2}^\perp$.

We obtain similarly that any element $\x$ of $\CoZd{\G_1}{\G_2}$ which is of weight less than
$\min(d_2,d_1^T)$ will belong to $\CoXd{\G_1}{\G_2}^\perp$. Therefore any element in 
the union of $\CoXd{\G_1}{\G_2} \setminus 
\CoZd{\G_1}{\G_2}^\perp$  and of  $\CoZd{\G_1}{\G_2} \setminus 
\CoXd{\G_1}{\G_2}^\perp$ should have weight at least $\min(d_1,d_2,d_1^T,d_2^T)$.
\end{proof}

\subsection{An upper bound on the minimum distance}

\begin{lemma}\label{lem:upper_bound}
Let $i$ belong to $\{1,2\}$. 
Assume that $d_{3- i} \neq \infty$. Then $\dQ \leq d_i$. If $d_{3- i}^T \neq \infty$ then
$\dQ \leq d_i^T$.
\end{lemma}

\begin{proof}{}
We will prove here that if $d_2 \neq \infty$ then $\dQ \leq d_1$. 
 The other inequalities are proved
in a similar fashion.

As before, $\G_1=\Tan{V_1}{C_1}{E_1}$ and $\G_2=\Tan{V_2}{C_2}{E_2}$
denote the two Tanner graphs used to define the quantum code
$\QC{\G_1}{\G_2}$ and $\Co_1$ and $\Co_2$ are the binary codes of
Tanner graphs $\G_1$ and $\G_2$ respectively.
 
 It will again be convenient to identify codewords with their
 supports, and we will allow ourselves the use of set operations
 $\cap,\subset$ and $\times$ on vectors and likewise we will use vector addition on sets.
 
Consider a codeword $\x_1$ of $\Co_1$ of weight $d_1$. 
Since $d_2 \neq \infty$, $\Co_2$ is not reduced to the zero word, and
therefore $\Co_2^\perp$ does not contain the whole of $\{0,1\}^{|V_2|}$.
Therefore there exists an element $y$ of $V_2$ such that $\{y\}$
 does not belong to $\Co_2^\perp$.
Let $\x = \x_1 \times \{y\}$. Our objective is to show that $\x$ is a weight $d_1$
codeword of $\CoXd{\G_1}{\G_2}$ that is not in $\CoZd{\G_1}{\G_2}^\perp$.

The vector (set) $\x$ 
is clearly an element of $\CoXd{\G_1}{\G_2}$ since the only check
nodes of the Tanner graph $\GXd{\G_1}{\G_2}$ incident to $\x$ are
nodes of the form $(c_1,y)$ with $c_1$ incident to $\x_1$ in $\G_1$.

We now assume that $\x$ is also an element of
$\CoZd{\G_1}{\G_2}^\perp$
and work towards a contradiction.

$\CoZd{\G_1}{\G_2}^\perp$ is generated by the rows $\h_Z(v_1,c_2)$ of
the parity-check matrix $\HH_Z$, with $(v_1,c_2)$ ranging over $V_1
\times C_2$: viewed as a set, $\h_Z(v_1,c_2)$ is the neighborhood in
$\G_1\times \G_2$ of vertex $(v_1,c_2)$. Recall from the definition of
$\G_1\times \G_2$ that $\h_Z(v_1,c_2)$ is the union of all nodes
$(v_1,v_2)$ such that $v_2$ is adjacent to $c_2$ in $\G_2$ and all
nodes $(c_1,c_2)$
such that $c_1$ is adjacent to $v_1$ in $\G_1$.

If $\x$ is also a codeword of $\CoZd{\G_1}{\G_2}^\perp$, then there exists 
a subset $U$ of $V_1 \times C_2$ such that
$$
\x = \bigoplus_{(v_1,c_2) \in U} \h_Z(v_1,c_2).
$$
Notice now that since $\x \subset \x_1 \times V_2$, we can write
\begin{eqnarray}
\x &= &\bigoplus_{(v_1,c_2) \in U} \h_Z(v_1,c_2) \cap (\x_1 \times V_2 )\\
& = & \bigoplus_{\substack{
    (x,c_2)\\ (x,c_2) \in U, x \in \x_1}}
\h_Z(x,c_2) \cap (\x_1 \times V_2) \label{eq:V},
\end{eqnarray}
since $ \h_Z(v_1,c_2) \cap (\x_1 \times V_2 ) = \emptyset$ for $v_1 \notin \x_1$.
For an element $x$ in $\x_1$ we denote by
$$
A(x) = \bigoplus_{\substack{c_2\\(x,c_2) \in U}} \h_Z(x,c_2) \cap (\x_1 \times V_2)
$$
Notice that 
\begin{equation}
\label{eq:A'}
A(x) \subset \{x\} \times V_2
\end{equation}
By combining this remark with $\x_1\times \{y\} = \x = \bigoplus_{x \in \x_1} A(x)$,
we obtain that for any $x$ in $\x_1$,
$$
\{(x,y)\} = A(x).
$$
On the other hand, we notice that
$$A(x)=\bigoplus_{\substack{c_2\\(x,c_2) \in U}} \bigoplus_{v_2 \sim
  c_2} \{(x,v_2)\}$$ 
where by $v_2 \sim c_2$ we mean that $v_2$ is adjacent to $c_2$ in $\G_2$.
This implies
$$
\{(x,y)\} =  \bigoplus_{\substack{c_2\\(x,c_2) \in U}} \bigoplus_{v_2 \sim c_2} \{(x,v_2)\}
$$
This in turn implies that
$$
\{y\} = \bigoplus_{\substack{c_2\\(x,c_2) \in U}} \bigoplus_{v_2 \sim c_2} \{v_2\}
$$
which means that 
$\{y\}$ is in  $C_2^\perp$. This contradicts the assumption made on $y$.
\end{proof}

Lemmas~\ref{lem:lower_bound} and \ref{lem:upper_bound} together prove
Theorem~\ref{th:main}. Theorem~\ref{th:main_intro} is obtained by
applying Theorem~\ref{th:main} with $\G_1=\G_2$ the Tanner graph of an
$[n,k,d]$ code.

\section{Comparison with other constructions of quantum codes based on binary linear codes}

Our construction can be viewed as a way of producing a quantum code from two
classical binary codes $\Co_1$ and $\Co_2$. The CSS construction achieves the same purpose
but requires that $\Co_2^\perp \subset \Co_1$. Using this construction directly to obtain quantum LDPC
codes is delicate due to the aforementioned orthogonality constraint. Our construction does not require
this constraint and gives a quantum LDPC code when $\Co_1$ and $\Co_2$ are classical LDPC 
codes. If we denote the length of $\Co_i$ by $n_i$, its dimension by $k_i$, its co-dimension $n_i-k_i$ by $r_i$, its minimum distance by $d_i$ and if we choose
a full rank parity-check matrix $\HH_i$ for it which describes the Tanner graph 
$\G_i$ used in the construction, then a straightforward application 
of the previous results leads to a quantum code $\QC{\G_1}{\G_2}$ with parameters 
$$[[n_1n_2+r_1r_2,k_1 k_2, \min(d_1,d_2)]].$$
Notice that $\CoXd{\G_1}{\G_2}$ has a parity-check matrix with the block form
$$
\HH_X = \begin{pmatrix}  \HH_1 \otimes \II_{n_2} & \II_{r_1} \otimes \HH_2^T\end{pmatrix}
$$
whereas $\CoZd{\G_1}{\G_2}$ has a parity-check matrix of the form
$$
\HH_Z = \begin{pmatrix}  \II_{n_1} \otimes \HH_2   &   \HH_1^T \otimes \II_{r_2}\end{pmatrix}
$$
where $\II_t$ stands for the $t \times t$ identity matrix. Notice that under this form, the property ensuring what we 
indeed define in this way a valid CSS code which is $\HH_X \HH_Z^T = 0$ can be verified directly
\begin{eqnarray*}
\HH_X . \HH_Z^T & = &\HH_1 \otimes \II_{n_2}. (\II_{n_1} \otimes \HH_2)^T + \II_{r_1} \otimes \HH_2^T.(\HH_1^T \otimes \II_{r_2})^T \\
& = & \HH_1 \otimes \HH_2^T + \HH_1 \otimes \HH_2^T \\
& = & 0.
\end{eqnarray*}

This construction displays some similarities with the generalized Shor code construction, see \cite{BC06b}, which produces with the
help of two binary linear  codes $\Co_1$ and $\Co_2$ a quantum code with parameters $[[n_1n_2,k_1 k_2, \min(d_1,d_2)]]$.
It is a CSS code like our construction and is associated to a couple $(\Co_X,\Co_Z)$ of binary codes satisfying
$\Co_X^\perp \subset \Co_Z$ defined by the following parity check matrices
$$
\HH_X = \HH_1 \otimes \II_{n_2}, \;\; \HH_Z = \GG_1 \otimes \HH_2.
$$
where $\GG_1$ is a generator matrix for $\Co_1$. Notice that our construction has the same dimension and minimum distance
as the generalized Shor  code construction but has an additional term $r_1 r_2$ in the length which compares favorably to Shor's construction.
Moreover, whereas our construction when applied to sparse matrices $\HH_i$'s yields sparse matrices $\HH_X$ and $\HH_Z$,
this is not the case in Shor's construction: $\HH_X$ stays sparse, however as soon as the minimum distance of $\Co_1$ is large,
this is not the case anymore for $\HH_Z$. Unlike our construction, the generalized Shor code construction is unable to yield quantum LDPC code families with non constant minimum distance.

\bibliographystyle{plain} 

\end{document}